\documentclass{aa}
\usepackage{graphicx}
\usepackage{footnote}
\usepackage{amssymb}
\begin{document}

\title{Radio continuum observations of the Leo Triplet at 2.64 GHz$^*$}

\author {
B. Nikiel-Wroczy\'nski\inst{1}
\and M. Soida \inst{1} 
\and M. Urbanik\inst{1}
\and M. We\.zgowiec\inst{2}
\and R. Beck\inst{3}
\and D. J. Bomans\inst{2}$^{,}$\inst{4}
\and B. Adebahr\inst{2}
}

\institute{Obserwatorium Astronomiczne Uniwersytetu
Jagiello\'nskiego, ul. Orla 171, 30-244 Krak\'ow, Poland
\and Astronomisches Institut, Ruhr-Universit\"at Bochum,
Universit\"atsstrasse 150, 44780 Bochum, Germany
\and Max-Planck-Institut f\"ur Radioastronomie, Auf dem
H\"ugel 69, 53121 Bonn, Germany
\and Research Department ``Plasmas with Complex Interactions'',
Ruhr-Universit\"at Bochum, Universit\"atsstrasse 150, 44780 Bochum, Germany}

\offprints{B. Nikiel-Wroczy\'nski}
\mail{iwan@oa.uj.edu.pl\\
$^*$Based on observations with the 100-m telescope of the 
MPIfR (Max-Planck-Institut f\"ur Radioastronomie) at Effelsberg}

\date{Received 11 January 2013/ Accepted 15 March 2013}

\titlerunning{2.64 GHz mapping of the Leo Triplet}
\authorrunning{B. Nikiel-Wroczy\'nski et al.}

\abstract
{The Leo Triplet group of galaxies is best known for the impressive bridges of 
neutral gas that connect its members. One of the bridges forms a large
tidal tail extending eastwards from NGC\,3628 that hosts several 
$\ion{H}{i}$ plumes and carries the material from this galaxy to
the intergalactic space.
}
{

The magnetic fields of the member galaxies NGC\,3628 and NGC\,3627
show morphological peculiarities, suggesting that interactions
within the group may be caused by stripping of the magnetic field. This process
could supply the intergalactic space with magnetised material, a scenario
considered as a possible source of intergalactic
magnetic fields (as seen eg. in the ``Taffy'' pairs of galaxies). Additionally, the plumes are likely to be the tidal
dwarf galaxy candidates.
}
{
We performed radio continuum mapping observations at 2.64\,GHz using the 100-m Effelsberg radio telescope.
We obtained total power and polarised intensity maps of the
Triplet. These maps were analysed together with the archive data, and
the magnetic field strength (as well as the magnetic energy density) was
estimated.
}
{Extended emission was not detected either in the total power or the polarised
intensity maps. We obtained upper limits of the magnetic field strength and
the energy density of the magnetic field in the Triplet. We detected emission from the easternmost clump
and determined the strength of its magnetic field. In addition, we 
measured integrated fluxes of the member galaxies at 2.64\,GHz and estimated their total
magnetic field strengths.
}
{We found that the tidal tail hosts a tidal dwarf galaxy candidate that possesses
a detectable magnetic field with a non-zero ordered component. 
Extended radio continuum emission, if present, is weaker than the reached confusion
limit. The total magnetic field strength does 
not exceed 2.8\,$\rm{\mu}$G and the ordered component is lower than 1.6\,$\rm{\mu}$G.
}
\maketitle

\section{Introduction}

Galaxy groups are known to contain large reservoirs of intergalactic gas 
(Trinchieri et al. \cite{trinchnew}). The intergalactic matter 
is a subject to violent tidal interactions (Hickson \cite{hickson}) caused by gravitational forces 
of the member galaxies. Studies of $\ion{H}{i}$ morphologies revealed that 
there is a variety of objects that contain outflows visible in the neutral hydrogen 
line (e.g. Williams et al. \cite{hcg44}, Yun et al. \cite{m81}, Stierwalt et al. \cite{hinew}).
There are two different types of intergalactic $\ion{H}{i}$ emission: ``wells'' and ``streamers''.
Which of these is present depends on the compactness of the 
system. In compact groups, the gravitational 
forces prevent the gas from escaping from the intergalactic space between the galaxies, and the emitting medium
is usually contained in a potential well between them 
(e.g. HCG\,44,  Williams et al. \cite{hcg44}, or the ``Taffy'' pairs of galaxies, 
Condon et al. \cite{tafold}, \cite{tafnew}; Drzazga et al. \cite{evol}). Conversely, loose groups are sometimes
called ``streamers'', because the emitting gas -- now weakly bound to the group members --
forms bridges and tails that extend far into the intergalactic space, 
like in the M81/M82 group (Yun et al. \cite{m81}).

The magnetic fields can play a significant role in the evolution and dynamics of galaxy systems, because
the energy densities of the thermal and magnetic components may be comparable. 
Not much is known about the magnetic fields in galaxy groups, however. Several studies
were carried out in compact groups, especially in Stephan's Quintet. Xu
et al. (\cite{xu}) detected intergalactic nonthermal emission (thus an intra-group magnetic 
field) in that group. Our own studies (Nikiel-Wroczy\'nski et al. in prep.) revealed
intergalactic polarised emission between its member galaxies. The magnetic field energy density
was determined to be approximately 5$\times10^{-12}$ erg/$\mathrm{cm^3}$ -- comparable 
to the thermal energy density estimated from X-ray observations (Trinchieri et al. 
\cite{trinchold}). Additionally, we found that the strength of the magnetic field of a tidal 
dwarf galaxy candidate SQ-B is comparable to that of normal spiral galaxies.

The origin of the intergalactic magnetic fields is lively debated and a variety of physical 
processes was proposed (e.g. reviews by Rees (\cite{rees}) and Stone (\cite{stone})). Usually, two scenarios
are suggested: Kronberg et al. (\cite{kron}) proposed dwarf irregular galaxies with strong winds 
that expell magnetic fields into intergalactic space (which was subsequently supported by numerical simulations by 
Siejkowski et al. \cite{hub}). Because these dwarfs are not always present in galaxy
groups, the scenario of the production of the magnetic field in spiral galaxies and interaction-driven
supply to the intergalactic space (as in the Antennae galaxies, Chy\.zy \& Beck \cite{antennae}) is 
another possibility. However, insufficient observational data make it impossible to distinguish whether one
of these mechanisms is dominating or another explanation is needed. Likewise, little is known if
 magnetic fields in tidal dwarf galaxy (TDG) candidates are a common phenomenon. The properties and 
strength of these fields are scarcely explored.
Thus, deep studies of various galaxy groups, of both streamer and compact types, are needed.

The Leo Triplet, also known as Arp\,317 (Arp \cite{arp}), is an example of a nearby streamer group.
With its distance of 15.7\,Mpc it is among the nearest objects of this 
type. The group consists of 
three tidally interacting spiral galaxies, NGC\,3623, NGC\,3627, and NGC\,3628. The galaxies are connected
by bridges of neutral gas (Stierwalt et al. \cite{hinew}, Haynes et al. \cite{arecibo}). The most
prominent one is the tidal tail that extends eastwards from NGC\,3628, which is visible also in the infrared at
60 and 100 microns (Hughes et al. \cite{hughes}).

Since NGC 3627 possesses a perturbed magnetic field (see Soida et al. \cite{oldsoida}, \cite{midsoida}
for details), the question arises
if the field is being stripped from galaxies into the intergalactic 
medium. As a possible explanation of the unusual
morphology of NGC\,3627, including its magnetic field, a collision with 
a dwarf galaxy was recently suggested by
We\.zgowiec et al. (\cite{3627dwarf}). The tidal tail of the Leo Triplet
is known to host several $\ion{H}{i}$ clumps, spatially
coincident with optical traces of star formation reported by Chromey et al. (\cite{chrom}), 
indicating that they might be TDG candidates. This could lead to the conclusion that
both the intergalactic and interstellar matter of spiral galaxies in the Leo Triplet galaxy group 
can be influenced by dwarf objects via ram-pressure/collision
heating and magnetic field enhancement. Therefore, studying TDG 
candidates becomes an important part of the
studies of galaxy groups. 

The proximity of the Triplet and its angular size of nearly $1\degr$ mean that it is best observed 
with a single-dish telescope. Such observations are characterised by high sensitivity to extended
structures. This allows searching for traces of magnetic field in 
$\ion{H}{i}$ clumps -- which constitute TDG candidates because they exhibit traces of stellar formation.
Therefore we performed a deep mapping of the Triplet using the 
100-m Effelsberg radio telescope at 2.64\,GHz. Our results are presented and discussed below.

\section{Observations}
\label{obs}
We mapped the Triplet in June 2012
with the 100-m Effelsberg radio telescope. Observations were performed
around the frequency of 2.64\,GHz, using a single-beam receiver
installed in the secondary focus of the telescope. The receiver has eight channels
and total bandwidth of 80\,MHz. The first 
channel (central frequency 2604\,MHz) was dropped due to the radio frequency interference (RFI).
We performed 12 coverages with a size of $80\arcmin\times80\arcmin$ each, scanned alternatively along 
R.A. and Dec.
We used a scanning velocity of $120\arcmin\slash$min. and a grid size of $2\arcmin$.
The data reduction was performed using the NOD2 data analysis system.
The maps were averaged and combined to reduce the scanning effects using the basket-weaving method 
(Emerson \& Gr\"ave \cite{bweav}), yielding final Stokes I, Q and U maps.
 During the observations we used
the radio source 3C286 to establish the flux density scale. The flux of 
the calibrator was taken from Mantovani et al. (\cite{fluxscale}). 
The polarisation was calibrated using 3C286 as well. We assumed
no circular polarisation (Stokes V signal = 0). The polarised fraction from
our observations is equal to 11.2$\pm 0.1\%$ and the polarisation position angle
is 33$\degr \pm 4\degr$ -- consistent with the values given in the
aforementioned article. The instrumental polarisation does not exceed 1$\%$.
The Astronomical Image
Processing System (AIPS) was used to produce the distribution of the polarised
intensity and the polarisation angle.
The resolution of the final maps is $4\farcm5$ (half--power beam width).

The uncertainties of the integrated flux densities include a $5\%$ uncertainty
of the flux scale determination.
The term ``apparent polarisation B--vectors'' used in this paper is
defined as the observed polarisation E--vectors direction rotated by $90\degr$, 
uncorrected for the Faraday rotation. Because the foreground Faraday rotation 
is relatively low (20$\degr$--30$\degr$ based on the data presented by Taylor et al. \cite{rm}),
these vectors constitute a reasonable approximation of the sky--projected 
magnetic field.

\section{Results}
Table~\ref{params} presents the properties of the detected sources that belong
to the Leo Triplet. Details on the total and polarised emission can be found in 
Sects.~\ref{TP} and \ref{PI}.
\begin{table*}[htbp]
\caption{Properties of the emission from the Leo Triplet sources at 2.64\,GHz}
\label{params}
\begin{center}
\begin{tabular}[]{rccccccc}
\hline
\hline
NGC& TP [mJy]& PI [mJy]&Pol. fract. [$\%$]& Mean pol. ang. [$\degr$]&$\alpha^{1}$
& Tot. magn. field [$\mu$G]\\
\hline
3623 & 19  $\pm$  2$^{2}$ & 1.3& 7 & 10 & n$\slash$a      & 3.5 $\pm$ 1\\
3627 & 299 $\pm$ 14 & 15 & 5 & 25 & 0.69 $\pm$ 0.14 &  11 $\pm$ 1\\
3628 & 364 $\pm$ 17 & 18 & 5 & 98 & 0.59 $\pm$ 0.15 &   9 $\pm$ 1.5\\
TDGc & 5.6 $\pm$ 1.0& 1.0&20 & 35 & 0.7  $\pm$ 0.3  & 3.3 $\pm$ 0.5\\
\hline
\end{tabular}
\end{center}
$^1$\footnotesize{
Calculated between our 2.64\,GHz observations and 1.4\,GHz data from Condon et al. (\cite{20cmcat})
}\\
$^{2}$\footnotesize{
Background sources not subtracted (see Sect. \ref{intflux} for details)
}
\end{table*}

\subsection{Total power emission}
\label{TP}

\begin{figure*}[ht!]
\sidecaption
  \resizebox{12cm}{!}{\includegraphics{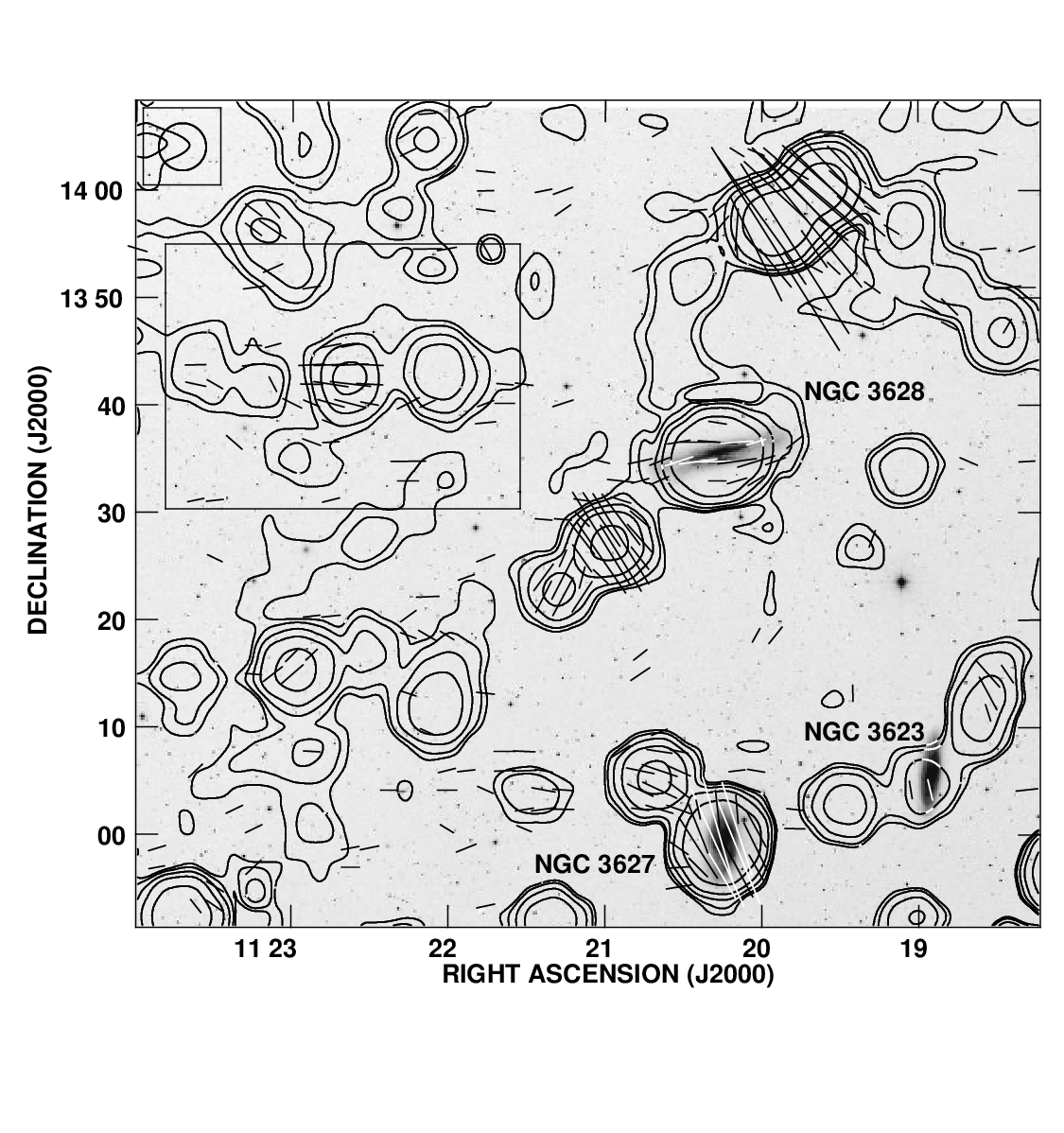}}
 \caption{ 
Contour map of the total power emission of the Leo Triplet at 2.64\,GHz.
Superimposed are the apparent B-vectors of polarised intensity
overlaid upon the blue image from DSS. The contour levels are
3,5,10,20,50, and $100\times 1.0\,\rm{mJy/beam}$. A polarisation
vector of 1$\arcmin$  corresponds to a polarised intensity 
of $1$\,mJy/beam. The beam size is $4\farcm5$.
The rectangular frame represents the area covered by Fig.~\ref{nvsseff}.
\newline
}
\label{tp11cm}  
\end{figure*}

Figure~\ref{tp11cm} shows the distribution of the total power (TP) emission
superimposed with apparent B-vectors of polarised intensity, obtained at the
center frequency of 2644\,MHz with the 100-m Effelsberg radiotelescope.
The r.m.s. noise level of the total power map is $1.0$\, mJy/beam.
The field of view is dominated by strong background sources. However, the 
highest signal comes from NGC\,3628. The total intensity of that galaxy is 
equal to $364 \pm 17$\, mJy. Owing to the high signal level, emission in the 
vicinity of NGC\,3628 is affected by the sidelobes of the telescope beam. Interferometric
observations (Dumke \& Krause \cite{dumke}) showed that the angular size of the 
galactic halo at 4.86\,GHz is around $3\arcmin$. A similar scaleheight has been reported at 1.49\,GHz (Reuter et al.
\cite{reuter}). Therefore the 2.64\,GHz emission should not exceed the main lobe of 
the beam and is not affected by its first negative sidelobe.

The second-most luminous object is the barred spiral NGC\,3627, with an integrated flux
of $299 \pm 14$\, mJy. The third member of the Triplet,
NGC\,3623, is also visible, but is significantly weaker with an integrated
flux of $19 \pm 2$\,mJy. It should be noted here that the large beam
of our observations causes flux contamination by the background sources at
$\mathrm{R.A._{2000}} = 11^{\rm{h}} 19^{\rm{m}} 01^{\rm{s}}$, $\mathrm{Dec_{2000}} = +13^{\circ}
01' 47''$ and $\mathrm{R.A._{2000}} = 11^{\rm{h}} 18^{\rm{m}} 57^{\rm{s}}$, $\mathrm{Dec_{2000}} = +13^{\circ}
04' 08''$. An attempt to estimate the real value at 2.46\,GHz is described in Sect.~\ref{intflux}.

Most of the emission comes from (background) sources that are not associated
with the Leo Triplet. There are nearly no signs of radio continuum emission from $\ion{H}{i}$ tails,
apart from a weak source coincident with an $\ion{H}{i}$ clump located at
$\mathrm{R.A._{2000}} = 11^{\rm{h}} 23^{\rm{m}} 11^{\rm{s}}$, $\mathrm{Dec_{2000}} = +13^{\circ}
42' 30''$ (flux of 5.6 $\pm$ 1.0\, mJy; it is marked with an arrow in Figs.~\ref{pi11cm} and \ref{nvsseff}). 
This source could be a tidal dwarf galaxy (TDG)
candidate; details can be found in Sect.~\ref{TDG}.

There is a tail-like structure north of NGC\,3628 (at $\mathrm{R.A._{2000}} = 11^{\rm{h}} 20^{\rm{m}} 20^{\rm{s}}$, $\mathrm{Dec_{2000}} = +13^{\circ}
45' 00''$), but a comparison with the NRAO VLA
Sky Survey (NVSS; Condon et al. \cite{nvss}) shows that this probably results from smoothing of two 
point sources with the large beam
of the telescope. Two luminous sources visible eastwards from NGC\,3628 (at $\mathrm{R.A._{2000}} = 11^{\rm{h}} 22^{\rm{m}} 00^{\rm{s}}$, $\mathrm{Dec_{2000}} = +13^{\circ}
42' 30''$ and $\mathrm{R.A._{2000}} = 11^{\rm{h}} 22^{\rm{m}} 40^{\rm{s}}$, $\mathrm{Dec_{2000}} = +13^{\circ}
42' 20''$ ) are not connected to the $\ion{H}{i}$ tail either.\\

\subsection{Polarised intensity}
\label{PI}
\begin{figure*}[ht!]
\sidecaption
   \resizebox{12cm}{!}{\includegraphics{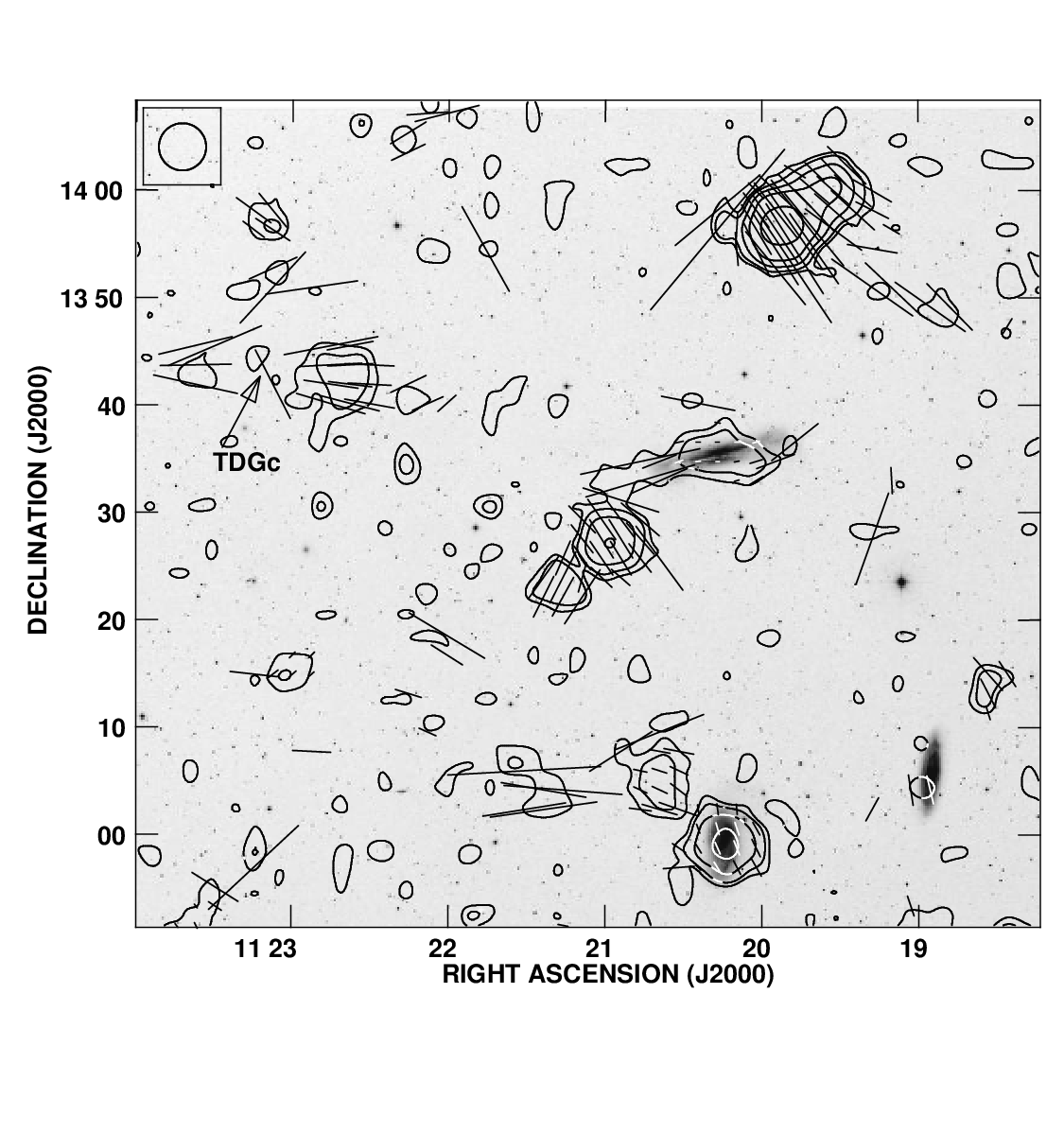}}
 \caption{ 
Contour map of the polarised intensity of the Leo Triplet at 2.64\,GHz.
Superimposed are the apparent B-vectors of the polarisation degree
overlaid upon the blue image from DSS. The contour levels are
3,5,10,20 and 50 $\times 0.35\,\rm{mJy/}$beam. A polarisation vector of 1$\arcmin$  
corresponds to a polarisation degree 
of $2\%$. The TDG candidate is marked with an arrow.
The beam size is $4\farcm5$.
\newline
}
\label{pi11cm}  
\end{figure*}

Figure~\ref{pi11cm} presents the distribution of the polarised intensity (PI) with
apparent B-vectors of the polarisation degree. As mentioned in Sect.~\ref{obs}, the expected 
amount of the Faraday rotation is approximately 20--30$\degr$. The r.m.s. noise level is equal to
0.35\, mJy/beam.
Similarly to the total power emission map, 
the polarised intensity distribution shows mostly background (point) sources that are not related to 
the object of study.
All three main galaxies were detected, with mean polarisation fraction
for NGC\,3623, NGC\,3627, and NGC\,3628 equal to $7\%$, $5\%$, and $5\%$.
The only extended structure is the disk of NGC\,3628. The PI vectors at 2.64\,GHz are
oriented along the galactic plane, resembling the structure previously reported at
4.86\,GHz by Soida (\cite{newsoida}).

There is also a marginal detection for the TDG candidate at the 2.5 $\sigma$ level.
Its polarisation degree reaches about 20$\%$. However, because the detection is very close to the noise level,
the real polarisation degree can be significantly different from the value we derived (it is presumably lower).

\section{Discussion}

\subsection{Integrated flux densities of the member galaxies}
\label{intflux}
The radio spectrum of the member galaxies has already been investigated
in several articles, allowing us to compare the flux values from
our study with previous works. Table~\ref{fluxes} presents a comparison of our data
with results from de Jong (\cite{dejong}), Kazes et al. (\cite{kazes}),
and Pfleiderer et al. (\cite{pflei}), who all studied
radio-luminous galaxies at 2.64\,GHz. For NGC\,3627 and NGC\,3628 we obtained similar
values; the better quality of the data resulted in lower measurement uncertainties.
For NGC\,3623 the data were much more scarce than for the other galaxies. 
Our value of 19 $\pm$ 2\, mJy is consistent with the one given by de Jong (\cite{dejong});
however, because of the background source contamination mentioned in Sect.~\ref{TP},
the real value is very likely lower. To estimate the real flux value, we calculated the
2.64\,GHz flux of the confusing sources using 1.4\,GHz data from the NVSS (Condon et al. \cite{nvss}) and assumed 
spectral index of 0.7. The integrated flux corrected for the background sources is equal to
6 $\pm$ 3 mJy.\\

To derive the spectral indices for the galaxies, we used data from Condon
et al. (\cite{20cmcat}). Our results can be found in Table \ref{params}. The value of 0.69$\pm$0.14 
derived for NGC\,3627 agrees well with the mean spectral index given by Soida et. al
\cite{oldsoida} (0.64$\pm$0.04). For NGC\,3628, our measurement (0.59
$\pm$0.15) and the mean spectral index from Dumke et al. \cite{dumke3628} (0.67$\pm$0.02) are
also agree well. For NGC\,3623, the contamination
prevents us from giving reliable estimate of its spectral index; however, using the flux value after
subtracting the background sources, we obtained $\alpha=$0.75 $\pm$ 0.2 -- which is within uncertainties 
consistent with the other two galaxies.

\begin{table*}[htbp]
\caption{Integrated fluxes (in mJy) for the member galaxies at 2.64\,GHz. References 
to the original publications can be found in the text.}
\label{fluxes}
\begin{center}
\begin{tabular}[]{rrrrrl}
\hline
\hline
NGC& this study& de Jong& Kazes&Pfleiderer\\
\hline
3623 & 6 $\pm$ 3$^{1}$ & $\leq$40&no data&no data\\
3627 & 299 $\pm$ 14 & 359 $\pm$ 28&206 $\pm$ 41& 310 $\pm$ 50\\
3628 & 364 $\pm$ 17& 280 $\pm$ 22&240 $\pm$ 48&410 $\pm$ 70\\
\hline
\end{tabular}
\end{center}
$^{1}$\footnotesize{
Background sources subtracted (see Sect. \ref{intflux} for details)
}
\end{table*}

\subsection{Constraints on the magnetic field strength of the Leo Triplet}
\label{field}

To calculate the magnetic field strength, we assumed energy
equipartition between the cosmic ray (CR) and the magnetic field energies and 
followed the procedure outlined by Beck \& Krause (\cite{bfeld}).
Because large-scale magnetised outflows were not detected, only upper constraints
for the magnetic field strength can be estimated. For the $\ion{H}{i}$ tail, we
adopted cylindrical symmetry and a diameter of at least 25\,kpc based on images 
presented by Stierwalt et al. (\cite{hinew}) and physical dimensions calculated from the 
NASA Extragalactic Database (NED). The extragalactic emission is typically characterised by a rather 
steep spectrum, with $\alpha > 1.0$ (eg. Pacholczyk \cite{pachol}, Chy\.zy \& Beck \cite{antennae}).
However, this refers to the CR electrons, not to the protons, which have higher energies than
the electrons and lose their energy more slowly (Beck \& Krause \cite{bfeld}). Moreover, higher
losses of the CR electrons result in an increase of the proton-to-electron ratio $K_0$. On the other hand,
in dense star-forming regions, interactions between CR protons and the nuclei of the IGM gas
may result in the generation of the secondary electrons 
(eg. Dennison \cite{dennison}, Ensslin et al. \cite{ensli}), for which it is yet unknown if the equipartition
formula is applicable. Recent investigations by Lacki \& Beck (\cite{lacki}) show 
that it is still valid for starburst galaxies, but nothing is known for the intergalactic fields.
However, because the generation of secondary electrons would be a rather extreme scenario, we 
assumed that the $K_0$ value is equal to 100. The magnetic field strength is rather weakly bound to the
proton-to-electron ratio; their dependence is given by a power-law function, and for the Leo Triplet, a change of
the $K_0$ value by an order of magnitude results in adjusting the magnetic field strength by not more than a 
factor of two. The nonthermal spectral index was estimated to be 1.1.\\

To obtain the constraints, we used the r.m.s. values
for the TP and PI emission as the upper limits for the nonthermal intensity.
Using the parameters mentioned above, we obtained a total magnetic field 
strength of $\lesssim 2.8\,\rm{\mu}$G, an ordered field 
component of $\lesssim 1.6\,\rm{\mu}$G and a magnetic field energy density of
$\lesssim 3.4 \times 10^{-13}$ erg/$\mathrm{cm^3}$. Comparing these values with those
obtained for Stephan's Quintet (Nikiel-Wroczy\'nski et al. in prep.) and the ``Taffy'' galaxy pairs
(Condon et al. \cite{tafnew}, Drzazga et al. \cite{evol}), the
large-scale magnetic field -- if present -- has an energy density lower by
approximately two orders of magnitude than in
the Quintet and the first Taffy pair, and about an order of magnitude lower than in the second Taffy pair. 
It is possible, however, that the magnetic field energy in the Leo Triplet
is not in equipartition with the CR energy.\\

To make the picture more complete, we calculated the magnetic field strength also for the member galaxies.
For NGC\,3627 and NGC\,3628 we used the spectral index values derived in Sect.~\ref{intflux} and a
pathlength of 2--4 kpc and 8--12 kpc, respectively. For NGC\,3623 we decided to use our estimate of 0.75 $\pm$ 0.2 (calculated
after subtracting the background sources); the pathlength was adopted as 6--8 kpc.
The magnetic fields of the first two galaxies (see Table \ref{params}) are relatively strong, although similar to the median 
value for the galaxies in the field, 9$\pm$1.3 $\mu$G (Niklas \cite{fields}). A slightly higher result obtained
for NGC\,3627 might be due the aforementioned collision with a dwarf galaxy (We\.zgowiec et al. \cite{3627dwarf}). The magnetic field of
NGC\,3623 is weaker; however, this galaxy is smaller than the other two, which may explain the difference.
It should be noted here that the values presented in this paragraph should be used only to visualize the
magnitude of the magnetic field, not the exact value, due to the uncertainties in the parameter estimation.
Especially for NGC\,3623, the magnetic field strength estimate is influenced by the flux
(and spectral index) estimate error due to the background sources.

\subsection{Possible tidal dwarf galaxy?}
\label{TDG}

\begin{figure}[ht!]
 \begin{center}
 \resizebox{8cm}{!}{\includegraphics{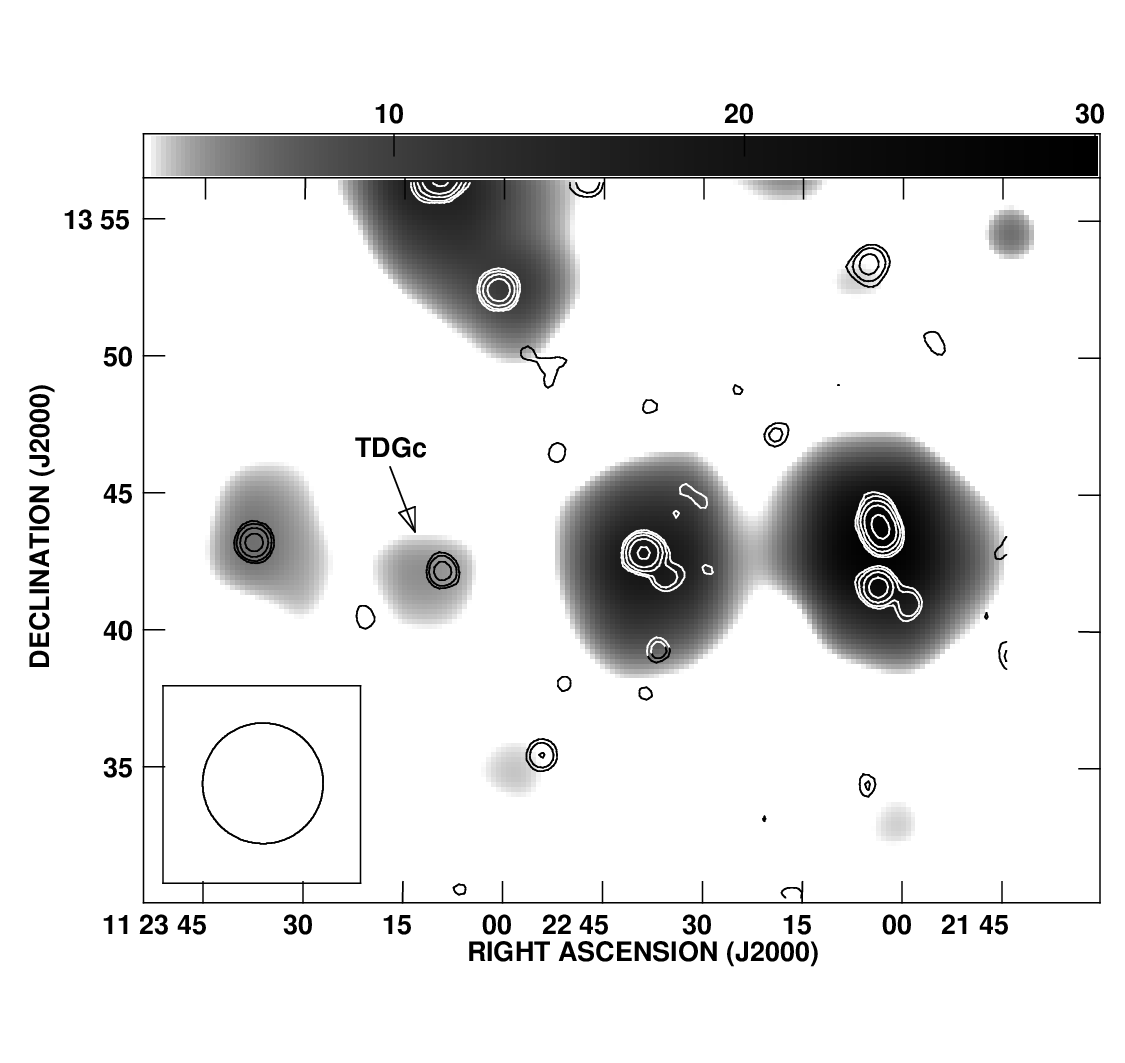}}
 \end{center}
 \caption{ 
Contours of the total power emission taken from 
the NVSS overlaid on a greyscale Effelsberg map of the TDG candidate
at 2.64\,GHz.
The contour levels are
3,5,10,20, and $50 \times 0.45\,\rm{mJy/beam}$.
The TDG candidate is marked with an arrow.
The beam size of the contour
map is $45 \arcsec$. The faintest structures in the greyscale
are at the 3 $\sigma$ level. 
The beam ellipse in the picture represents the $4\farcm5$ 
Effelsberg beam.
}
\label{nvsseff}  
\end{figure}

As mentioned in Sect.~\ref{TP}, only one of the radio continuum sources visible in the TP
map belongs to the $\ion{H}{i}$ tail. This source is spatially coincident
with the $\ion{H}{i}$ clump reported by Stierwalt et al. (\cite{hinew}) that is also associated 
with faint emission in the optical domain (Chromey et al. \cite{chrom})
and with the local maximum in the IRAS infrared data (Hughes et al. \cite{hughes}).
The coincidence of the recent (optical) and past (infrared) stellar formation and 
neutral gas indicates that this object
is a TDG candidate. Such objects were first proposed by Zwicky (\cite{zwicky}).
Since the detection of a TDG candidate in the Antennae galaxies (Mirabel
et al. \cite{mirabel}), such sources have frequently been reported in interacting
systems (see e.g. Kaviraj et al. \cite{kaviraj}).\\
We investigated available VLA Faint Images of the Radio Sky at Twenty-Centimeters 
(FIRST; Becker et al. \cite{first}) and NVSS (Condon et al. 
\cite{nvss}) data and found that the source was not detected by the FIRST survey,
but there is a detection in the NVSS (which is more sensitive to extended emission than FIRST).
The total flux at 1.4\,GHz is equal to 8.4 $\pm$ 0.5 \,mJy. Fig.~\ref{nvsseff} presents 
contours from the NVSS overlaid on the 2.64\,GHz data, clearly showing the 1.4\,GHz counterpart for our
detection (marked with an arrow). The faint source near $\mathrm{R.A._{2000}} = 11^{\rm{h}} 23^{\rm{m}} 40^{\rm{s}}$, $\mathrm{Dec_{2000}} = +13^{\circ}
43' 30''$  is a background source. \\
We calculated the spectral index using our 2.64\,GHz and NVSS data and obtained a
value of 0.7 $\pm$ 0.3. This
could indicate that the radio continuum emission is coming from a
recent star--formation period, supporting our claim for a TDG candidate. The size of
the TDG candidate was assumed to be 4--6\, kpc, because Chromey et al. (\cite{chrom}) gave 5.1\,
kpc as the size of the stellar clump. Using the same method as in Sect.~\ref{field}, we obtained a 
total magnetic field strength $ \approx 3.3\pm0.5\,\rm{\mu}$G and a magnetic field energy density
$\approx 4.5 \pm 1.5 \times 10^{-13}$ erg/$\mathrm{cm^3}$. These values are somewhat lower than for
the TDG candidate found in Stephan's Quintet (Nikiel-Wroczy\'nski et al. in prep.), but still reasonable 
for a large-scale magnetic field: Niklas (\cite{fields}) derived $\rm{B_{TOT}}= 9 \pm 3\,\rm{\mu}$G as mean for 
the normal-sized spiral galaxies -- a value about three times higher.
The clump was marginally detected in the PI map.
If we attribute this detection to the TDG candidate, it gives
a polarisation degree of about 20$\%$ (see Sect.~\ref{PI})
and yields a strength of the ordered
magnetic field of $\approx 1.5\pm0.3\,\rm{\mu}$G.

\section{Summary and conclusions}

We observed the Leo Triplet of galaxies using the
100-m Effelsberg radio telescope at a frequency of 2.64\,GHz. We obtained
maps of total power (TP) emission and polarised intensity (PI). These maps were 
analysed together with the available archive data, yielding the following results:
\begin{itemize}
 \item[--] Although the high sensitivity of our observations did not allow a direct detection of
 the $\ion{H}{i}$ bridges in the radio
 continuum at 2.64\,GHz (neither in the TP nor in the PI map),
 we were able to calculate upper limits for the magnetic field in the intergalactic space of the Leo Triplet.
 Assuming equipartition between the magnetic field and the CRs, the total field is not stronger than $2.8\,\mu$G, 
 with an ordered component at most of $1.6\,\mu$G.
 \item[--]  NGC\,3623, NGC\,3627, and NGC\,3628 are 
 visible both in the TP and the PI map. The easternmost
 $\ion{H}{i}$ clump -- visible also in the optical and infrared domains -- was also detected.
 \item[--] The $\ion{H}{i}$ clump detected in radio continuum could be an example of a tidal dwarf galaxy because there are traces
 of recent star formation and infrared emission in its position.
 The strength of its magnetic field is about $3.3\pm0.5\,\mu$G and the magnetic 
 field energy density reaches $4.5 \pm 1.5 \times 10^{-13}$ erg/$\mathrm{cm^3}$. These
 values -- although lower than estimated for another known TDG candidate, the SQ-B source
 of Stephan's Quintet -- are high enough to indicate a large-scale magnetic
 field. The ordered component is about $1.5 \pm 0.3\,\mu$G.
 \item[--] The total fluxes for the main galaxies are equal to 19 $\pm$ 2 \,mJy for NGC\,3623
 (6 $\pm$ 3 \,mJy after subtraction of background sources),
 299 $\pm$ 15 \,mJy for NGC\,3627, and 364 $\pm$ 18 \,mJy for NGC\,3628. The mean 
 polarisation degree is 7, 5, and 5$\%$, respectively.
 \end{itemize}
 The planned analysis of existing X-ray data for the Leo Triplet spiral 
 galaxies will allow a comparison of the estimated intergalactic magnetic 
 fields with the thermal properties of both galaxies and the group 
 gaseous medium. This will provide important constraints on evolutionary 
 models of galaxy groups.

\begin{acknowledgements}
We would like to thank an anonymous referee for helpful comments
and suggestions. 
BNW and MS are indebted to the whole staff of the radio telescope
Effelsberg for all the help and guidance during the observations. 
We thank Marita Krause from the MPIfR Bonn for valuable comments.
We acknowledge the usage of the HyperLeda database (http://leda.univ-lyon1.fr) and 
the NASA/IPAC Extragalactic Database (NED), which is operated by the Jet Propulsion Laboratory,
California Institute of Technology, under contract with the National Aeronautics and Space Administration. 
This research has made use of NASA's Astrophysics Data System.
This research has been supported
by the scientific grant from the National Science Centre (NCN), decision no. DEC-2011/03/B/ST9/01859
and by the Dean of the Faculty of Physics, Astronomy and Applied Computer Sciences of the Jagiellonian University,
decision no. DSC/0708/2012.
RB and DJB acknowledge support by the DFG SFB\,591 `Universal 
Behaviour of non-equilibrium plasmas' and DFG FOR\,1254, `Magnetisation of Interstellar
and Intergalactic Media`.
\end{acknowledgements}

\end{document}